\def\3he{$^3$He}
\def\4he{$^4$He}
\def\7li{$^7$Li}
\def\la{\mathrel{\mathpalette\fun <}}
\def\ga{\mathrel{\mathpalette\fun >}}
\def\fun#1#2{\lower3.6pt\vbox{\baselineskip0pt\lineskip.9pt
  \ialign{$\mathsurround=0pt#1\hfil##\hfil$\crcr#2\crcr\sim\crcr}}}
\def\beq#1\eeq{\begin{equation}#1\end{equation}}

\def\Yp{Y$_{\rm P}$}
\def\hii{H\thinspace{$\scriptstyle{\rm II}$}}
\def\hi{H\thinspace{$\scriptstyle{\rm I}$}}
\newcommand{\Deln}{\ensuremath{\Delta N_\nu}}
\newcommand{\nnu}{\ensuremath{N_\nu}}

\documentclass{iaus}
\usepackage{graphicx}

\title[IAUS 265.~~Primordial Nucleosynthesis] 
{Primordial Nucleosynthesis After WMAP}

\author[Gary Steigman]   
{Gary Steigman
}

\affiliation{ Departments of Physics and Astronomy, Center for Cosmology and 
Astro-Particle Physics, The Ohio State University\\ 191 West Woodruff Avenue, 
Columbus, OH 432120, USA \\ email: {\tt steigman@mps.ohio-state.edu}}

\pubyear{2009}
\volume{265}  
\pagerange{1--9}
\jname{Chemical Abundances in the Universe: Connecting First Stars to Planets}
\editors{K. Cunha, M. Spite \& B. Barbuy, eds.}
\begin{document}

\maketitle

\begin{abstract}
During its early evolution, the hot, dense Universe provided a laboratory 
for probing fundamental physics at high energies.  By studying the relics 
from those early epochs, such as the light elements synthesized during 
primordial nucleosynthesis when the Universe was only a few minutes 
old, and the relic, cosmic microwave photons, last scattered when the 
protons, alphas, and electrons (re)combined some 400 thousand years 
later, the evolution of the Universe may be used to test the standard 
models of cosmology and particle physics and to set constraints on 
proposals of physics beyond these standard models.			
\end{abstract}

\firstsection 
\section{Introduction}

Primordial, or Big Bang, Nucleosynthesis (BBN) provides a key probe of 
the physics and early evolution of the Universe, as do observations of 
the cosmic microwave background (CMB) radiation.  These probes offer 
windows on the early Universe at two widely separated epochs: BBN when 
the Universe was only $\sim 20$ minutes old and the CMB some 400 thousand 
years later.  BBN and the CMB provide complementary tests of the consistency 
of the standard, hot big bang cosmology and offer observational tests 
of its quantitative predictions.  For a recent review, see \cite[Steigman 
(2007)]{steigman07}; for a comparison between the predictions of BBN and 
the CMB, see \cite[Simha \& Steigman (2008)]{simha08}.

The standard models of cosmology, and of standard, big bang nucleosynthesis 
(SBBN), employ the general theory of relativity (GR) to describe an expanding 
Universe, filled with radiation (including three flavors of light neutrinos) 
and matter (including non-baryonic dark matter).  For our purposes here, 
the presence or not of dark energy is irrelevant since dark energy (or a 
cosmological constant) plays no role in the physics of the early Universe 
which concerns us here.  For BBN, the relic abundances of D, \3he, \4he, 
and \7li are predicted as a function of two cosmological parameters: the 
baryon density parameter $\eta_{\rm B}$ and the expansion rate parameter 
$S \equiv H'/H$, where $H$ is the standard model value of the Hubble 
parameter and $S \neq 1$ allows for a large class of non-standard models 
of particle physics and/or cosmology.  For SBBN it is assumed that the 
Hubble parameter assumes its standard model value ($S = 1$), so that the 
relic abundances depend on only one cosmological parameter, $\eta_{\rm B}$.

Here, to assess the current status of the standard models of particle 
physics and cosmology, we'll ask if the BBN-predicted and observationally 
inferred relic abundances of the light nuclides agree and, if the 
BBN-determined values of $\eta_{\rm B}$ and $H$ are consistent with the 
values inferred from independent (non-BBN) cosmological observations, 
including those of the CMB and of the Large Scale Structure (LSS) 
observed in the Universe.

\section{Defining The Cosmological Parameters}

{\underline{\it Baryon Density Parameter}}  

In the relatively late, early Universe, at the time of BBN (and later), 
the only baryons present are the nucleons, the neutrons and protons.  
Hence, ``baryons" and ``nucleons" will be used interchangeably.  As the 
Universe expands, all densities decrease so, to define a parameter which 
provides a measure of the baryon/nucleon abundance, it is convenient 
(and conventional) to compare the baryon number density to the number 
density of CMB photons: $\eta_{\rm B} \equiv n_{\rm B}/n_{\gamma}$.  
Since $\eta_{\rm B}$ is very small, it is convenient to introduce 
$\eta_{10} \equiv 10^{10}\eta_{\rm B}$.  In terms of the baryon density 
parameter, the baryon {\it mass} density parameter $\Omega_{\rm B} \equiv 
\rho_{\rm B}/\rho_{\rm crit}$ may be written as $\Omega_{\rm B}h^{2} = 
\eta_{10}/274$, where the present value of the Hubble parameter is $H_{0} 
\equiv  100h$~km~s$^{-1}$~Mpc$^{-1}$ (\cite{steigman06b}).

The annihilation of relic electron-positron pairs  in the early Universe, 
when $T \la m_{e}c^{2}$, produces ``extra" photons which are thermalized 
and become part of the CMB observed today.  Since BBN occurs after $e^{\pm}$ 
annihilation is complete and, since baryons are conserved, the numbers 
of baryons and CMB photons in every comoving volume in the Universe are 
(should be) unchanged from BBN to recombination to the present epoch 
(N$_{\rm B}/$N$_{\gamma} = \eta_{\rm B} =$~constant).  As a result, the 
value of $\eta_{\rm B}$ inferred from the comparison of the BBN predictions 
with the abundance observations should agree with the value inferred from 
the CMB (supplemented by observations of LSS).

{\underline{\it Expansion Rate Parameter}}  

For the standard cosmology the expansion rate (the Hubble parameter) 
during the early evolution of the Universe is determined solely by the 
mass/energy density of the Universe: $H^{2} \propto G\rho$, where $G$ is 
Newton's gravitational constant and $\rho$ is the energy density which, 
during the early evolution of the Universe, is dominated by ``radiation" 
(e.g., massless or relativistic particles).

In the presence of non-standard physics and/or cosmology (\eg modifications 
to GR or, to the standard model particle content leading to $\rho 
\rightarrow \rho' \neq \rho$),
\beq
S^{2} = (H'/H)^{2} = G'\rho'/G\rho \equiv 1 + 7\Delta{\rm N}_{\nu}/43,
\eeq
where
\beq
\Delta{\rm N}_{\nu} \equiv (\rho' - \rho)/\rho_{\nu}.
\eeq
\Deln~parameterizes any difference from the standard model, early Universe 
predicted energy density, normalized to the contribution from one additional 
light neutrino.  For SBBN, N$_{\nu} = 3$ and $S = 1$ (\Deln~= 0).  However, 
since any departures from the standard models may arise from new particle 
physics and/or new cosmology, \Deln~does not necessarily count additional 
flavors of neutrinos and, indeed, \nnu~may be $<  3$ or $> 3$ (i.e., $S < 1$ 
or $S > 1$).  \nnu~(or \Deln) is simply a convenient way to parameterize a 
non-standard, early Universe expansion rate.  For example, if the particle 
content of the standard model of particle physics ($\rho' = \rho$) is adopted,
\beq
G'/G = 1 + 7\Delta{\rm N}_{\nu}/43,
\eeq
the BBN or CMB determined values of \nnu~can constrain any deviations 
between the early Universe magnitude of Newton's constant and the value 
measured terrestrially today.

\section{Primordial Nucleosynthesis}

{\underline{\it Standard BBN}}  
\label{sbbn}

For standard BBN (SBBN) the relic abundances of the light nuclides produced 
during primordial nucleosynthesis depend only on the baryon density parameter 
$\eta_{\rm B}$.  Over a limited range in $\eta_{\rm B} \equiv 10^{10}\eta_{10}  
\approx 6 \pm 1$, the SBBN-predicted abundances (\cite{ks04,steigman07}) 
depend on the baryon density parameter as, (D/H)$_{\rm P} \propto \eta_{10}
^{-1.6}$, (\3he/H)$_{\rm P} \propto \eta_{10}^{-0.6}$, and (\7li/H)$_{\rm P} 
\propto \eta_{10}^{2.0}$.  A good fit to the SBBN-predicted \4he mass fraction 
is \Yp~$= 0.2485 \pm 0.0005 + 0.0016(\eta_{10} - 6)$.  The SBBN-predicted 
relic abundances of D and \7li are most sensitive to the baryon density 
parameter, while those of \3he and \4he are less (for the latter, much less) 
sensitive to $\eta_{\rm B}$.

{\underline{\it Non-Standard BBN}}  
\label{bbn}

For non-standard BBN ($S \neq 1$, \nnu~$\neq 3$), it is the primoridal 
abundance of \4he which provides the most sensitive probe.  According 
to \cite{ks04}, for $0.85 \la S \la 1.15$ ($1.3 \la$~N$_{\nu} \la 5.0$) and 
$5 \la \eta_{10} \la 7$, (D/H)$_{\rm P} \propto \eta_{\rm D}^{-1.6}$, 
where $\eta_{\rm D} \equiv \eta_{10} - 6(S - 1)$ and (\7li/H)$_{\rm P} 
\propto \eta_{\rm Li}^{2.0}$, where $\eta_{\rm Li} \equiv \eta_{10} - 
3(S - 1)$.  In contrast to the relatively weak dependences on $S$ of the 
D and \7li abundances, a good fit to the primordial \4he abundance is 
\Yp~$= 0.2482 \pm 0.0006 + 0.0016(\eta_{\rm He} - 6)$, where $\eta_{\rm 
He} \equiv \eta_{10} + 100(S - 1)$.  While deuterium (or \7li) probes 
the baryon density parameter, \4he is sensitive to \nnu.

{\underline{\it Deuterium: The Baryometer Of Choice}}  

\begin{figure}[b]
\begin{center}
 \includegraphics[width=3.4in]{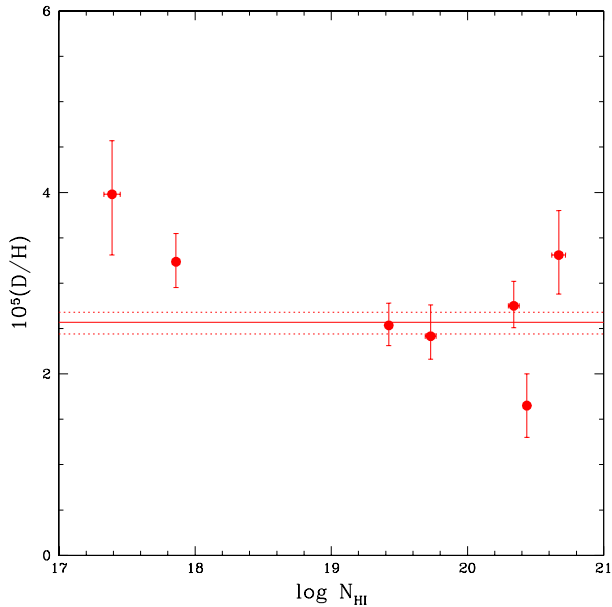} 
 \caption{The deuterium abundances, $y_{\rm D} \equiv 10^{5}$(D/H), 
 inferred from observations of high-redshift, low-metallicity QSOALS, 
 as a function of the log of the corresponding \hi~column densities.  
 The solid line is at the value of the weighted mean of the seven 
 $y_{\rm D}$ values, $\langle y_{\rm D} \rangle = 2.7$, and the dotted 
 lines show the estimated uncertainty, $\sigma(y_{\rm D}) = \pm$~0.2.}
   \label{dvsh}
\end{center}
\end{figure}

Of the BBN-synthesized light nuclei, deuterium is the baryometer of choice.  
One key reason is that  the post-BBN evolution of deuterium is simple: 
as gas is cycled through stars, deuterium is destroyed.  As a result, 
the deuterium abundance measured anywhere in the Universe, at any time 
during its evolution, provides a {\bf lower limit} to the primordial D 
abundance.  In particular, if the D abundance is measured in high-redshift, 
low-metallicity systems where post-BBN stellar synthesis has been minimal, 
the observed abundances should approach the primordial value (``deuterium 
plateau": as $Z \rightarrow 0$, (D/H)$_{\rm OBS} \rightarrow$~(D/H)$_{\rm 
P}$).  Furthermore, as already noted, the relic D abundance is sensitive 
to the the baryon density parameter.  For SBBN, a $\sim 10\%$ determination 
of (D/H)$_{\rm P}$ results in a $\sim 6\%$ constraint on $\eta_{10}$.  That's 
the good news.  The bad news is that high precision, high spectral resolution 
observations of D at high-redshifts and low-metallicities (e.g., in high$-z$, 
low$-Z$ QSO Absorption Line Systems (QSOALS; see, e.g., \cite{pettini} and 
earlier reference therein) are difficult, requiring significant observing 
time on large telescopes, equipped with high resolution spectrographs.  
As a result, as shown in Figure \ref{dvsh}, at present there are only seven, 
relatively reliable D abundance determinations.  

The weighted mean of the seven D abundances is $y_{\rm DP} \equiv 
10^{5}(D/H)_{\rm P} = 2.7$ (note that the weighted mean of log($y_{\rm DP})$ 
is 0.45, which corresponds to $y_{\rm DP} = 2.8$) but, as may be seen from 
the Figure \ref{dvsh}, only three of the seven abundances lie within $1 \sigma$ 
of the mean.  Indeed, the fit to the weighted mean of these seven data points 
has a $\chi^{2} = 18$ ($\chi^{2}/dof = 3$).  Either the quoted errors in the 
inferred D abundances are too small or, one or more of the determinations 
are wrong, perhaps contaminated by unidentified (and, therefore, uncorrected) 
systematic errors. In the absence of further evidence identifying the reason(s) 
for such a large dispersion, the best that can be done at present is to adopt 
the mean D abundance but to inflate the error in the mean in an attempt to 
account for the unexpectedly large dispersion among the D abundances 
(\cite{steigman07}).  
\beq
y_{\rm DP} \equiv 10^{5}({\rm D/H})_{\rm P} = 2.7 \pm 0.2.
\label{ydp}
\eeq
Note that the \cite{pettini} value of log($y_{\rm DP}) = 0.45 \pm 0.03$ 
corresponds to  $y_{\rm DP} = 2.8 \pm 0.2$, consistent, within the errors, 
with the weighted mean of the individual $y_{\rm D}$ values.  For quantitative 
comparisons, the value of $y_{\rm DP}$ from eq.~(\ref{ydp}) is adopted here.
 
For SBBN, this value of the primordial D abundance corresponds to 
$\eta_{10} = 6.0 \pm 0.3$ or, $\Omega_{\rm B}h^{2} = 0.022 \pm 0.001$ 
(\cite{steigman07}), a 5\% determination of the baryon density parameter.  
For comparison, if the \cite{pettini} value were adopted, a slightly lower 
(but consistent) value, $\eta_{10} = 5.8 \pm 0.3$ or, $\Omega_{\rm 
B}h^{2} = 0.021 \pm 0.001$, would be found.

{\underline{\it Non-BBN Determinations Of The Baryon Density Parameter: 
CMB And LSS}}  

The baryon density parameter determined by SBBN and the deuterium 
observations reflects the value of this parameter when the Universe 
is some $\sim 20$ minutes old.  According to standard model physics 
and cosmology, the value of this parameter should be unchanged some 
$\sim 400$ thousand years later, at recombination (and, at present, 
some $\sim 14$ Gyr later).   The comparison between the BBN-determined 
baryon density parameter and that inferred from observations of the 
CMB (see, e.g., \cite{wmap} and further references therein) and of LSS 
provides a test of the standard models of particle physics and cosmology 
(\cite{steigman07}).  According to \cite{simha08}, the combination of the 
CMB plus LSS data results in $\eta_{10} = 6.1 \pm 0.2$ or, $\Omega_{\rm 
B}h^{2} = 0.022 \pm 0.001$.  More recent results from the WMAP team 
(\cite{dunkley09, komatsu09}) and others (\cite{sanchez09}) suggest a 
slightly higher value of $\eta_{10} = 6.22 \pm 0.16$ or, $\Omega_{\rm 
B}h^{2} = 0.0227 \pm 0.0006$.  Within the uncertainties, $\eta_{\rm 
B}$(BBN)~$ \approx \eta_{\rm B}$(CMB/LSS); the number of baryons 
(nucleons) in a comoving volume of the Universe is unchanged between 
BBN and recombination (as it should be for the standard model).

\section{Testing SBBN}

Having found agreement between $\eta_{\rm B}$(BBN) and $\eta_{\rm 
B}$(CMB/LSS), we still need to test the consistency of SBBN.  That is, 
do the abundances of the other light nuclides (\3he, \4he, \7li) predicted 
by SBBN, using the D-determined value of $\eta_{\rm B}$, agree with 
their observationally inferred primordial abundances?

\begin{figure}[b]
\begin{center}
 \includegraphics[width=3.4in]{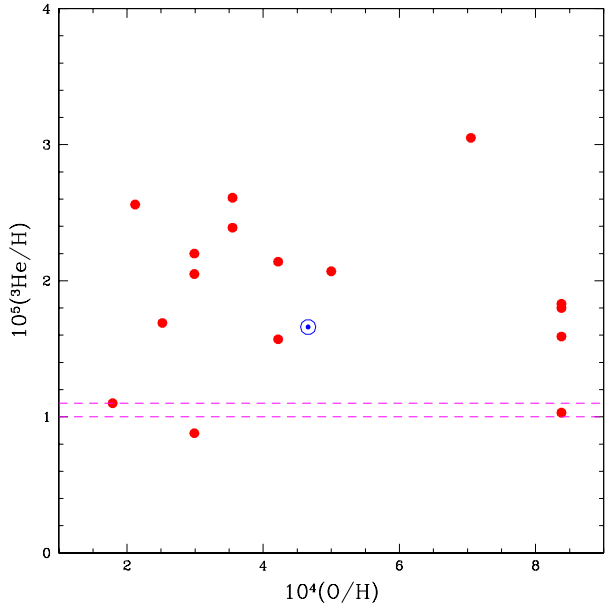} 
 \caption{The \3he abundances, $y_{3} \equiv 10^{5}$(\3he/H), 
 from observations of Galactic \hii~regions (\cite{bania}), as a 
 function of the \hii~region oxygen abundances.  The solar symbol 
 is the pre-solar nebula \3he abundance (\cite{gg}).  The dashed 
 lines show the $\pm 1\sigma$ range around the SBBN-predicted 
 primordial abundance, $y_{\rm 3P} = 1.05 \pm 0.05$.}
   \label{3he}
\end{center}
\end{figure}

{\underline{\it Helium-3:}}  

In contrast to deuterium, the post-BBN evolution of \3he is complex.  
When gas is cycled through stars, D is burned to \3he, any prestellar 
\3he (prestellar D + \3he) is burned away in the hot interiors (of all 
but the least massive stars), but preserved in their cooler, outer 
layers and, new \3he is synthesized via stellar nucleosynthesis 
in the interiors of lower mass stars (e.g., \cite{iben}, \cite{rood}, 
\cite{rst99}).  Competition between destruction, preservation, and 
synthesis, complicates the process of using the observations to 
infer the primordial \3he abundance.  The data (see \cite{bania}) 
don't help.  Observations of \3he are limited to the solar system 
\cite{gg} and to \hii~regions in the Galaxy (see \cite{steigman06} and 
\cite{steigman07} for further discussion and references).  If, in the 
course of Galactic chemical evolution  there is a net increase in 
the abundance of \3he from its primordial value, the \3he abundance 
and metallicity should be correlated (and, the \3he abundance in the 
interstellar medium (ISM) of the Galaxy at present should exceed the 
presolar nebula abundance).  As may be seen in Figure \ref{3he}, no 
clear trend with metallicity is revealed and, while many of the 
observed \3he abundances do exceed the solar abundance, some don't.  
All that may be inferred from this data (in this author's opinion) 
is that the {\it lowest} \3heabundances observed are consistent with 
the SBBN-predicted primordial abundance and, the remaining \3he 
abundances suggest net production of \3he in the course of Galactic
chemical evolution.  D, \3he, and the CMB/LSS are in agreement with SBBN.

\begin{figure}[b]
\begin{center}
 \includegraphics[width=3.4in]{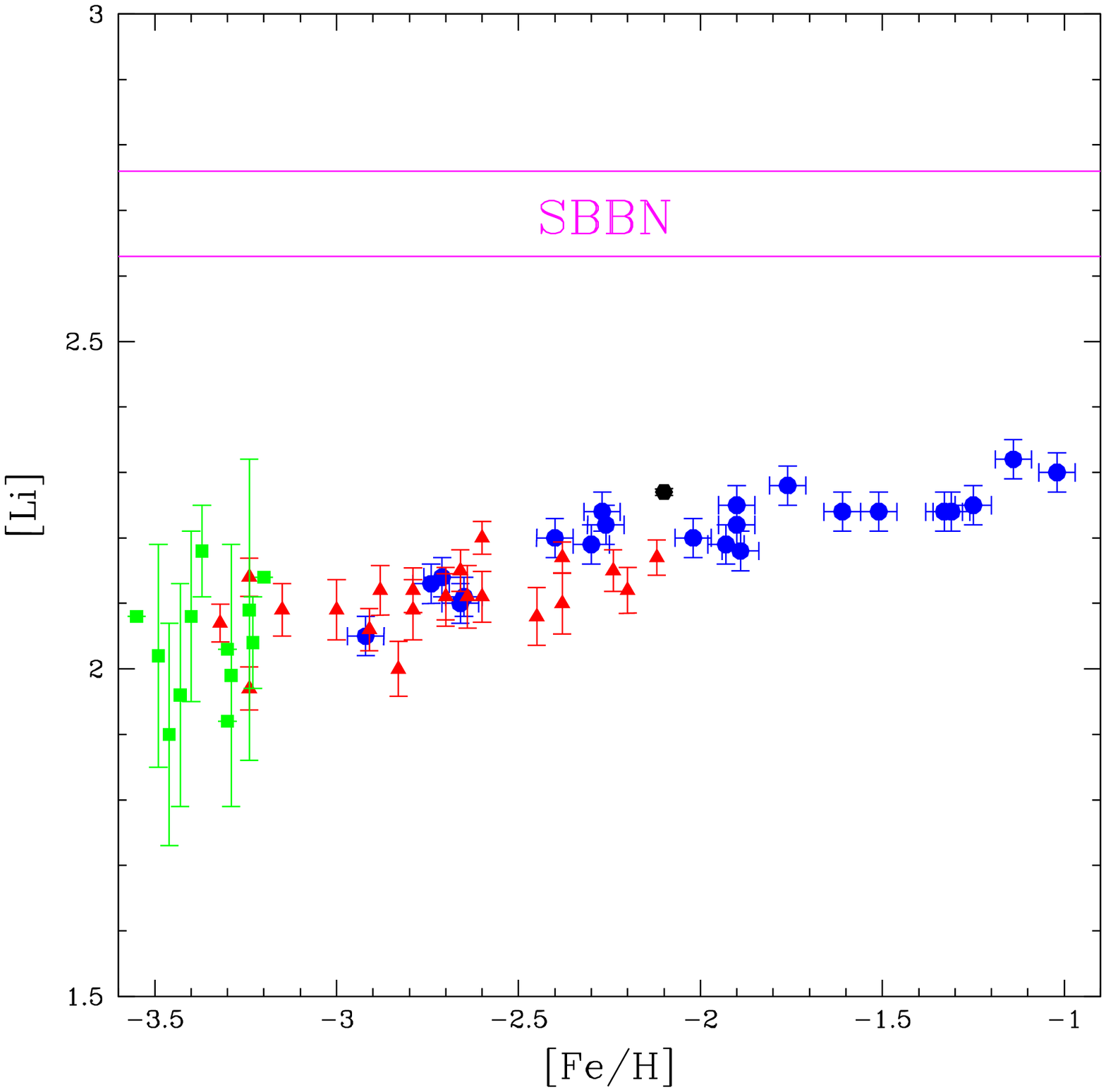} 
 \caption{The log of the \7li abundances, [Li]~$\equiv 12 + $log(\7li/H), 
 from observations of old, very metal-poor halo stars, as a function of 
 the stellar iron abundances.  Blue filled circles (\cite{asplund06}), red, 
 filled triangles (\cite{boesgaard05}), green filled squares (\cite{aoki09}). 
 The black filled circle (\cite{lind09}) is for the globular cluster NGC6397.
 The solid lines shows $\pm 1\sigma$ range around the SBBN-predicted 
 primordial abundance [Li]$_{\rm SBBN} = 2.70 \pm 0.06$.}
   \label{7li}
\end{center}
\end{figure}

{\underline{\it Lithium-7:}}  

\7li is a relatively fragile, weakly bound nuclide, easily destroyed at the 
high temperatures inside most stars.  However, as with \3he, some \7li may 
survive in the cooler, outer layers of stars and, stellar production and 
cosmic ray nucleosynthesis likely increase the post-BBN abundance of \7li.  
The data reveal that while lithium is depleted in many stars, the overall 
trend is a lithium abundance which increases with metallicity.  As the 
metallicity approaches zero (primordial), the \7li abundances are expected 
to plateau (the ``Spite plateau") at the primordial abundance.  In Figure 
\ref{7li} the lithium abundances, [Li]~$\equiv 12 + $log(\7li/H), are shown 
as a function of the iron abundances (on a log scale, normalized to the 
solar iron abundance) for the most metal-poor halo stars (and for the 
globular cluster NGC 6397).  Where is the Spite plateau?  The data in 
Figure~\ref{7li} (see, also, the contributions to these proceedings by 
Sbordone \etal\ and Melendez \etal) fail to reveal clear evidence for a 
plateau as [Fe/H] $\rightarrow 0$.  Even more disturbing is the fact that 
{\bf none} of the lithium abundances inferred from these observations of 
the oldest, most metal-poor, most nearly primordial stars in the Galaxy, 
come even close to the SBBN-predicted abundance.  The observed
abundances are too low by factors of $\sim 3 - 5$.  This gap seems
too wide to be closed by observational uncertainties.  Either this
conflict between the predictions of SBBN and the observations is
pointing to ``new" physics and/or cosmology or, our understanding
of the structure and evolution of the oldest, most metal-poor stars
in the Galaxy is seriously incomplete. 

{\underline{\it Helium-4:}}

After hydrogen, helium (\4he) is the most abundance element in the 
Universe.  In the post-BBN Universe, as gas cycles through stars, 
the helium abundance (the mass fraction of \4he, Y) increases from 
its primordial value \Yp.  While the correction for stellar produced 
\4he is uncertain, its contribution can be minimized by restricting 
attention to the most metal-poor sites, the low-metallicity, 
extragalactic \hii~regions (Blue Compact Galaxies).  The current 
status of the search for \Yp~is an object lesson in the difference 
between quantity and quality.  With more than $\sim 100$ helium 
abundance determinations, statistical uncertainties are very small: 
$\sigma($Y$_{\rm P})_{stat} \la 0.001$ (\cite{izotov}).  However, most 
analyses fail to deal adequately with the many identified (but often 
ignored) sources of systematic errors, whose values are estimated to 
be larger, $\sigma($Y$_{\rm P})_{syst} \ga 0.006$ (see \cite{steigman07} 
for a discussion of these and other related issues and, for further 
references).  Following \cite{steigman07} and \cite{simha08}, here we 
adopt \Yp~$=0.240 \pm 0.006$ as an estimate of the primordial \4he 
mass fraction.  For SBBN, this corresponds to  a very small value 
of the baryon density parameter, $\eta_{10}($\4he$) \la 3$ (note 
that the simple fitting formula for \Yp~in \S\ref{sbbn} is not valid 
for such a low helium abundance and the result here is from the 
full BBN code).  Such a low estimate of the baryon density parameter 
is clearly in conflict with the value determined by SBBN and D, 
which is otherwise consistent with the CMB/LSS determined value.  
How serious is this conflict?  That is, given the estimate of 
the error in the observationally determined value of \Yp, how 
bad is the disagreement?  For $\eta_{10}({\rm D}) = 6.0$, the 
SBBN-predicted primordial helium abundance is Y$_{\rm P} = 0.249$.  
The observationally-inferred abundance differs from the predicted 
abundance by only $\sim 1.5\sigma$, a not very serious disagreement.  
However, this tension between the predicted and observed helium 
abundances could be a hint of non-standard physics and/or cosmology.
 
\section{Extension Of the Standard Models: \nnu~$\neq 3$ ($S \neq 1$)}

\begin{figure}[b]
\vspace*{-0.5 cm}
\begin{center}
 \includegraphics[width=5.85in]{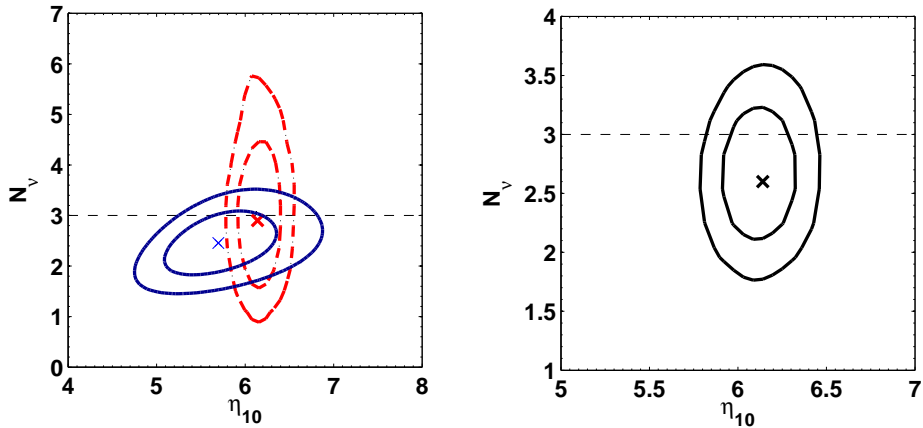} 
\vspace*{-7.2 cm}
 \caption{Left panel:  The 68\% and 95\% contours in the \nnu~vs. $\eta_{10}$
  plane (from \cite{simha08}) for BBN (D \& \4he) (solid) and for the 
  CMB/LSS (dashed).  The crosses indicate the best fit values (see the 
  text).  Right panel: The joint BBN/CMB/LSS contours; note the expanded 
  scales for \nnu~and $\eta_{10}$.
  }
   \label{nnuvseta}
\end{center}
\end{figure}

If, indeed, the tension between the observationally-inferred and SBBN-predicted 
primordial helium abundances is taken seriously, it is of interest to explore
non-standard models of particle physics and/or cosmology, with $S \neq 1$
(\nnu~$\neq 3$). For BBN and the D and \4he abundances adopted here, $\eta_{10} 
= 5.6 \pm 0.3$ and \nnu~$= 2.4 \pm 0.4$, confirming that the standard model 
(\nnu~= 3) is only $\sim 1.5\sigma$ away.  For the non-BBN CMB and LSS data, 
\cite{simha08} find $\eta_{10} = 6.14^{+0.16}_{-0.11}$ and \nnu~$= 2.9^{+1.0}_
{-0.8}$.  As shown in the left hand panel of Figure \ref{nnuvseta} (from 
\cite{simha08}), there is significant overlap between BBN and the CMB/LSS.  
BBN and the CMB agree and, at $\ga 68\%$ confidence, they are consistent 
with \nnu~= 3.  In the right hand panel of Figure \ref{nnuvseta} the likelihood 
\nnu~-- $\eta_{10}$ contours are shown for the combined BBN/CMB/LSS data 
(\nnu~$= 2.5 \pm 0.4$, $\eta_{10} = 6.1 \pm 0.1$).  

For the non-BBN constraints on the baryon density and expansion rate 
parameters, the BBN-inferred primordial abundances of D and \4he are 
$y_{\rm DP} = 2.5 \pm 0.3$ and Y$_{\rm P} = 0.247^{+0.013}_{-0.011}$, 
in good agreement, within the errors, with the adopted relic abundances.  
However, it must be noted that for the non-BBN identified values of 
$\eta_{10}$ and \nnu, the BBN-predicted lithium abundance, [Li]$_{\rm 
P} = 2.72^{+0.05}_{-0.06}$, remains in serious conflict with the 
observationally inferred value.  The lithium (\7li) problem persists.

\section{Conclusions}

The very good agreement between the values of the baryon density 
and universal expansion rate parameters determined by BBN, when 
the Universe was $\sim 20$ minutes old, and by the CMB/LSS, some 
$\sim 400$ thousand to 14 billion years later, leads to constraints on 
some extensions of the standard models of particle physics and cosmology.  

{\underline{\it Entropy Conservation?}}  

For example, the numbers of baryons and CMB photons in a comoving 
volume are related by the baryon density parameter, N$_{\rm B} = 
\eta_{\rm B}$N$_{\gamma}$.  In the standard model of particle physics, 
N$_{\rm B}$ is unchanged from the BBN to recombination and the 
present.  Comparing $\eta_{\rm B}$ at BBN and at recombination 
leads to the constraint: N$_{\gamma}$(CMB)/N$_{\gamma}$(BBN) 
$= 0.92 \pm 0.07$, limiting any post-BBN entropy production.

{\underline{\it ``Extra", Post-BBN Radiation Density?}}  

Since $\rho'_{\rm R}/\rho_{\rm R} = 1 + 7\Delta$N$_{\nu}/43$, the BBN 
and CMB constraints on \nnu~limit the radiation energy densities at 
these widely separated epochs.  In the absence of the creation of 
``new" radiation (e.g., by the late decay of a massive particle), 
\nnu(BBN) = \nnu(CMB).  Comparing \nnu~at BBN and at recombination 
constrains any possible difference between these values.  This 
comparison reveals that $0.94 \leq $~N$_{\nu} \leq 1.23$, consistent 
with {\bf no} extra, post-BBN radiation density.

{\underline{\it  Variation Of the Gravitational Constant?}}  

The BBN and CMB constraints on \nnu~also limit any difference between 
the magnitude of the gravitational constant at BBN or at recombination and 
that observed today terrestrially since $G'/G = 1 + 7\Delta$N$_{\nu}/43$.   
From \nnu~at BBN, $G_{\rm BBN}/G_{0} = 0.91 \pm 0.07$, while the 
CMB/LSS bound on \nnu~leads to an even tighter constraint, $G_{\rm 
CMB}/G_{0} = 0.99  \pm 0.12$. 

\section{Summary}  

For \nnu~$\approx 3$, BBN agrees with the observations of the CMB (and 
LSS and $H_{0}$), confirming the consistency of the standard models of 
particle physics and cosmology.  But, lithium remains a problem whose 
origin may lie with stellar depletion/dilution or with new particle physics 
and/or cosmology.  When BBN is combined with the CMB and LSS, 
interesting constraints on some non-standard models of particle physics 
and cosmology can be obtained.

\end{document}